\documentstyle[preprint,aps]{revtex}
\begin{document}
\draft
\tightenlines

\title{Microcanonical versus canonical ensemble of bosons in a 1D harmonic trap.}
\author{Muoi N. Tran}

\address
{Department of Physics and Astronomy, McMaster University\\
 Hamilton, Ont. L8S 4M1, Canada}
\date{\today}
\maketitle

\begin{abstract} 
We consider a fixed number of noninteracting bosons in a harmonic trap. The determination of the exact microcanonical ground state number fluctuation is a difficult enterprise.  There have been several theoretical attempts to solve this problem approximately, especially in 1D models where analytic results were found using some asymptotic formulae from number theory.  Here, we obtain the {\em exact} number fluctuation curves, and show that these exact curves are substantially different from the results presented in literature for small particle number and finite temperature.  Further, we show how these approximate microcanonical results may be derived using a canonical thermodynamics approach.  

\end{abstract}


\narrowtext 
\section{Introduction}
As is well known, the traditional approach of determining the fluctuation of single-particle occupation numbers fails for bosons as the temperature $T \rightarrow 0$.  It predicts that the fluctuation tends to infinity in this limit, rather than zero as described by a real system.  After the experimental success in producing Bose-Einstein condensation of trapped alkali-metal atoms, there is renewed interest in calculating the number fluctuation using either canonical or microcanonical ensemble approach \cite{grossmann1,grossmann2,gajda,navez,borrmann,holthaus,tran1}.  Most of these, however, use some forms of approximation to the real microcanonical fluctuation.  In \cite{navez} the authors define the so-called Maxwell's Demon ensemble, and calculate the microcanonical fluctuation for a 3D trap numerically.  In \cite{tran1} we formulated a method of calculating the {\em exact} microcanonical fluctuation using combinatorics for a perfectly isolated BEC with finite number of particles which, to our best knowledge, has not been presented before. 

In this report we focus on a system of bosons in an one-dimensional harmonic trap. The number fluctuation of this system was calculated by the authors in \cite{grossmann1}, using the well-known asymptotic formulae from number theory to approximate the exact expansion coefficients of the N-particle partition function. Their method yields results which were supposed to be {\em microcanonical}.  However, we shall show in this brief report that their approximate method yields the same results one obtains from canonical ensemble averaging, or a related formulation by Parvan {\it et al} \cite{parvan}.  At low temperatures and for finite number of particles ($N \leq 100$), these results are very different from the ones calculated using the {\em exact} combinatorics method. 

The layout of this paper is as follows.  In section \ref{microcanonical} we establish the notations and outline the method of calculating the {\em exact} ground state number fluctuation \cite{grossmann2,tran1,tran2}.  Section \ref{canonical} contains the main focus of this brief report in which we derive the results of \cite{grossmann1} using thermodynamics, and show that they agree with the ones obtained from canonical method as given by \cite{parvan}.  Finally, we compare the exact combinatorics results from section \ref{microcanonical} and the canonical results from section \ref{canonical}. 


\section{Microcanonical Formalism}
\label{microcanonical}
We consider an isolated system with discreet energy levels consisting of N particles in the ground state at zero temperature.  Given an excitation energy $E$, there can be one particle which absorbs all the energy and gets excited to an excited state, or two particles which share the available energy...etc.  The number of excited particles $N_{ex}$ can be somewhere between 1 and N, depending on the excitation quanta, such that the total number of particles $N$ remain constant.  This defines the microcanonical ground state number fluctuation.  Denote $\omega(E,N_{ex},N)$ to be the number of possible ways of distributing $E$ among {\em exactly} $N_{ex}$ particles, then the probability of exciting {\em exactly} $N_{ex}$ particles in a N-particle system at excitation energy $E$ is given by \cite{tran1,tran2}:
\begin{equation}
P(E,N_{ex},N) =  \frac{\omega(E,N_{ex}, N)}{\sum_{N_{ex}=1}^N \omega(E,N_{ex},N)}
\label{probability}
\end{equation} 

Expression (\ref{probability}) is normalized and has the following properties: 
\begin{eqnarray}
P(0,N_{ex},N) &=&  \delta_{0N_{ex}},\\
P(E,N_{ex},N) &=& 0 ~~~ N_{ex} > N.
\end{eqnarray} 

The moments of the probability distribution now read:
\begin{eqnarray}
\left< N_{ex}\right> &=& \sum_{N_{ex}=1}^{N} N_{ex} P(E,N_{ex},N)\\
\left< N_{ex}^2\right> &=& \sum_{N_{ex}=1}^{N} N_{ex}^2 P(E,N_{ex},N)
\end{eqnarray}
The number of particles in the ground state $N_0$ and the excited states $N_{ex}$ are related via $N=N_0+N_{ex}$. $N_0$ and $N_{ex}$ are allowed to vary while $N$ is fixed so that the number fluctuation from the ground state is given by:

\begin{eqnarray}
\left< (\delta N_0)^2\right> &=& \left< N_{ex}^2\right>  - \left< N_{ex}\right>^2 \nonumber \\
             &=& \left<N_0^2\right>  - \left< N_0\right>^2 
\label{deltanzero}
\end{eqnarray}
Expression (\ref{deltanzero}) is true in general regardless of the ensemble and statistics of the particles.  

Next, we specialize to a one-dimensional harmonic trap.  Let us put $\hbar \omega =1$, then the excitation energy from the ground state is $E=n$. The partition function of the system is well-known
\begin{eqnarray}
Z_1(\beta) &=& \frac{1}{1-x}\\
Z_N(\beta) &=& \prod_{j=1}^{N} \frac{1}{1-x^j}
\label{zn}
\end{eqnarray}
where $x=e^{-\beta}$.
This is just the generating function pertaining to the restricted linear partitions of an integer n in number theory \cite{rademacher}.
Expressing (\ref{zn}) in power series of $x$
\begin{equation}
Z_N(\beta)=\sum_{n=0}^ \infty\Omega(n,N)x^n
\label{znexpansion}
\end{equation}

The coefficients $\Omega(n,N)$ is the number of partitions of n which have at most $N$ summands.  For example, if $n=5$ and $N=4$, then $\Omega(5,4)=6$ since $5=5,1+4,2+3,1+1+3,1+2+2,1+1+1+2$. In connection to our system, $\Omega(n,N)$ is interpreted as the number of possible ways to distribute $n$ quanta among utmost $N$ particles.  Clearly,
\begin{equation}
\Omega(n,N)=\sum_{N_{ex}=1}^N \omega(E,N_{ex},N)
\end{equation}
While the coefficients $\Omega(n,N)$ may be found analytically by expanding $Z_N(\beta)$ in series, the exact determination of the microstate $\omega(E,N_{ex},N)$ is a difficult problem.  We may find $\omega(E,N_{ex},N)$ using combinatorics as described in detail in \cite{tran1}, or, alternatively, use the well-known identity \cite{grossmann2}
\begin{equation}
\omega(n,N_{ex},N)=\Omega(n,N_{ex})-\Omega(n,N_{ex}-1)
\label{recursive}
\end{equation}
Using (\ref{recursive}), (\ref{probability}) and (\ref{deltanzero}), the {\em exact microcanonical} ground state fluctuation may be determined. 
 

\section{Canonical formalism}
\label{canonical}
The ground state number fluctuation may in general be expressed in terms of the moments of occupation numbers as \cite{chase,parvan,tran2}:
\begin{equation}
\left<(\delta N_0)^2\right> = \sum_{k}\left(\left<n_k^2\right> - \left<n_k\right>^2\right)
\label{thermalfluct}
\end{equation}
The sum runs through all the allowed k values in the ground state defined at zero temperature.  For bosons, only $k=0$ state applies.  Canonically, the moments are given by:
\begin{eqnarray}
\label{avnkx}
\left<n_k\right> &=& \frac{1}{Z_N} \sum_{j=1}^N x^{j \epsilon_k}  Z_{N-j}(\beta)\\
\left<n_k^2\right> &=& \frac{1}{Z_N}\sum_{j=1}^{N}~\left[2j-1\right]~  x^{j \epsilon_k}Z_{N-j}(\beta)
\label{avnsqkx}
\end{eqnarray}

We can also obtain the canonical fluctuation using a thermodynamics approach \cite{tran1}.  Given the N-particle partition function (\ref{zn}), the entropy of the system reads
\begin{eqnarray}
S_N(\beta) &=& \beta \sum_{j=1}^N \frac{j}{[x^{-j} -1]} -
               \sum_{j=1}^N \ln [1-x^{j}]             ~\nonumber \\
           &=& \beta E - \sum_{j=1}^N \ln [1-x^{j}]
\label{ceentropy}
\end{eqnarray}
Note that the canonical multiplicity $\Omega(\beta,N)$ by definition is given by:
\begin{equation}
\Omega(\beta,N)=\exp[S_N(\beta)]~=x^{-E} Z_N(\beta)~ 
\label{thermalomega}
\end{equation} 
Compare (\ref{thermalomega}) with (\ref{znexpansion}), we see that the canonical multiplicity is a one-term approximation to the exact value of the expansion coefficients of the series (\ref{znexpansion}). 
To make a direct comparison with the results given in \cite{grossmann1}, we approximate the sums in (\ref{ceentropy}).  The first term gives the energy-temperature relationship:
\begin{equation}
E=n=\sum_{j=1}^N\frac{j}{[x^{-j} -1]}~\approx~\int_0 ^\infty \frac{j}{[x^{-j} -1]}~= \frac{\pi^2}{6}\frac{1}{\beta^2}
\label{Tton}
\end{equation}
To retain the N-dependence we use the Euler-Maclaurin sum formula in approximating the second term:
\begin{equation}
\sum_{j=1}^N \ln [1-x^{j}] \approx -\frac{\pi^2}{6}\frac{1}{\beta} + \frac{1}{\beta}x^N
\label{Ndependent}
\end{equation}
Using (\ref{recursive}) and (\ref{ceentropy}), the canonical probability of finding $N_{ex}$ particles in the excited states is given by:

\begin{eqnarray}
P(\beta,N_{ex},N) &=& \frac{\Omega(\beta,N_{ex}) - \Omega(\beta,N_{ex}-1)}{\Omega(\beta, N)}
                                                ~\nonumber \\
&=& \frac{1}{\Omega(\beta, N)}\frac{\partial~ \Omega(\beta,N_{ex})}{\partial N_{ex}}
						~\nonumber \\
&=& \frac{1}{exp[S(\beta,N)]}\frac{\partial~ exp[S(\beta,N_{ex})]}{\partial N_{ex}}
				 		~\nonumber \\
&=& e^{-\beta N_{ex}}~ exp \left[-\frac{1}{\beta}\left(e^{-\beta N_{ex}} - e^{-\beta N}\right)\right]
\label{canonicalprob}
\end{eqnarray}
Using (\ref{Tton}) and the transformations 

\begin{eqnarray*}
c=\sqrt{\frac{2}{3}} \pi,~~~ \chi_{r}=\frac{c~ r}{2\sqrt{n}}-ln(\sqrt{n})
\end{eqnarray*} 
then,

\begin{equation}
P(n,N_{ex},N) = \frac{1}{\sqrt{n}}\frac{exp\left[-\frac{2}{c}e^{-\chi_{N_{ex}}}-\chi_{N_{ex}}\right]}{exp\left[-\frac{2}{c}e^{-\chi_{N}}\right]}
\end{equation}
This is identical to expression (14) given in ref.~\cite{grossmann1} which the authors claim give the microcanonical fluctuation.  That the canonical method given by (\ref{thermalfluct}), (\ref{avnkx}), and (\ref{avnsqkx}) and the above method are equivalent is shown in Fig.~(\ref{figure1}) \cite{tran1}.  {\em Both yield the same canonical fluctuations}. 


\section{Microcanonical vs.~Canonical}
\label{mcvscn}
Fig.~(\ref{figure2}) compares the relative fluctuations for $N=$ 10, 50, 100 particles.  The respective curves of both ensembles tend to agree as $N$ and $n$ get large ({\it ie.}~thermodynamic limit), otherwise the results are very different.

It is clear from the discussion above that the method employed by the authors in ref.~\cite{grossmann1} yields {\em canonical} results.  Comparison between their method and the thermodynamics method  makes it clear on why this is so: both use what essentially is a saddle point approximation to the exact multiplicity $\Omega(n,N)$ ($p_N(m)$ in their notation) from the series (\ref{zn}).  Only when the multiplicity found exactly are the results microcanonical. This clearly makes it more difficult to incorporate interaction, and treat the true condensate fluctuations of weakly interacting isolated bosons microcanonically. 

The author thank R. K. Bhaduri and M. V. N. Murthy for fruitful discussions.  This work was supported by NSERC grant \# 199160-01 and NSERC scholarship \# PGSA-221708-1999.

\newpage
\begin{figure} 

\caption{Canonical ground state number fluctuation for $N=1000$ bosons in a 1D harmonic confinement.  The solid line shows the fluctuation using formulae (\ref{thermalfluct}), (\ref{avnkx}), and (\ref{avnsqkx}).  The diamonds were obtained using the thermodynamics method as described in section \ref{canonical}.}  
\label{figure1}
\vspace{7 mm}
\end{figure}
\begin{figure} 
\caption{Comparison between canonical and microcanonical relative fluctuations.  The critical temperature is given by $T_c=N/ln N$ with $\hbar \omega=1$ [1].  The inset shows a close-up of the low temperature part.} 
\label{figure2}
\vspace{7 mm}

\end{figure}


\newpage
\bigskip
\begin{references}

\bibitem{grossmann1}
S. Grossmann and M. Holthaus, Phys. Rev. {\bf E 54}, 3495 (1996).
\bibitem{grossmann2}
S. Grossmann and M. Holthaus, Phys. Rev. Lett {\bf 79}, 3557 (1997).
\bibitem{gajda}
M. Gajda and K. Rz\c{a}\.{z}ewski, Phys. Rev. Lett. {\bf 78}, 2686 (1997).
\bibitem{navez}
P. Navez, D. Bitouk, M. Gajda, A. Idziaszek, and K. Rz\c{a}\.{z}ewski, Phys. Rev. Lett. {\bf 79}, 1789 (1997).
\bibitem{borrmann}
P. Borrmann, J. Hartings, O. M\"{u}lken, and E. R. Hilf, Phys. Rev. {\bf A 60}, 1519 (1999). 
\bibitem{holthaus}
M. Holthaus and E. Kalinowski, Ann. of Phys. {\bf 276}, 321-360 (1999).
\bibitem{tran1}
M. N. Tran,{\it Ground state fluctuations in finite bose and fermi systems}. M.Sc. thesis, 2000. 
\bibitem{tran2}
M. N. Tran, M. V. N. Murthy, and R. K. Bhaduri, Phys. Rev. {\bf E 63}, 031105(2001).

\bibitem{rademacher}
H. Rademacher, {\it Topics in Analytic Number Theory} (Springer Verlag, Berlin, 1973).

\bibitem{chase}
K. C. Chase, A. Z. Mekjian, and L. Zamick, Eur. Phys. J. {\bf 8}, 281 (1999) 
\bibitem{parvan}
A. S. Parvan, V. D. Toneev, and M. Ploszajczak, Nucl. Phys. {\bf A 676}, 409 (2000)
\end{references}
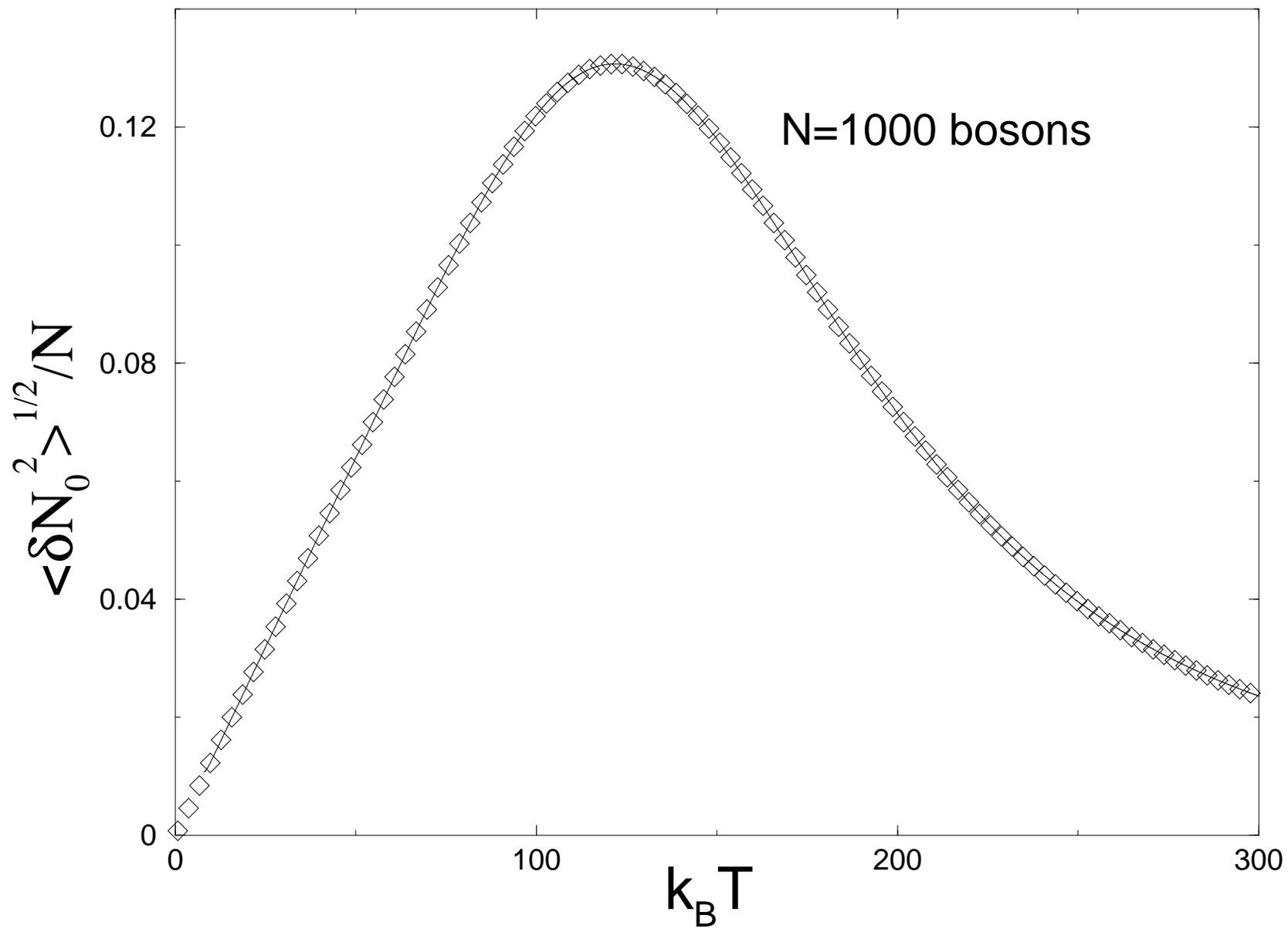
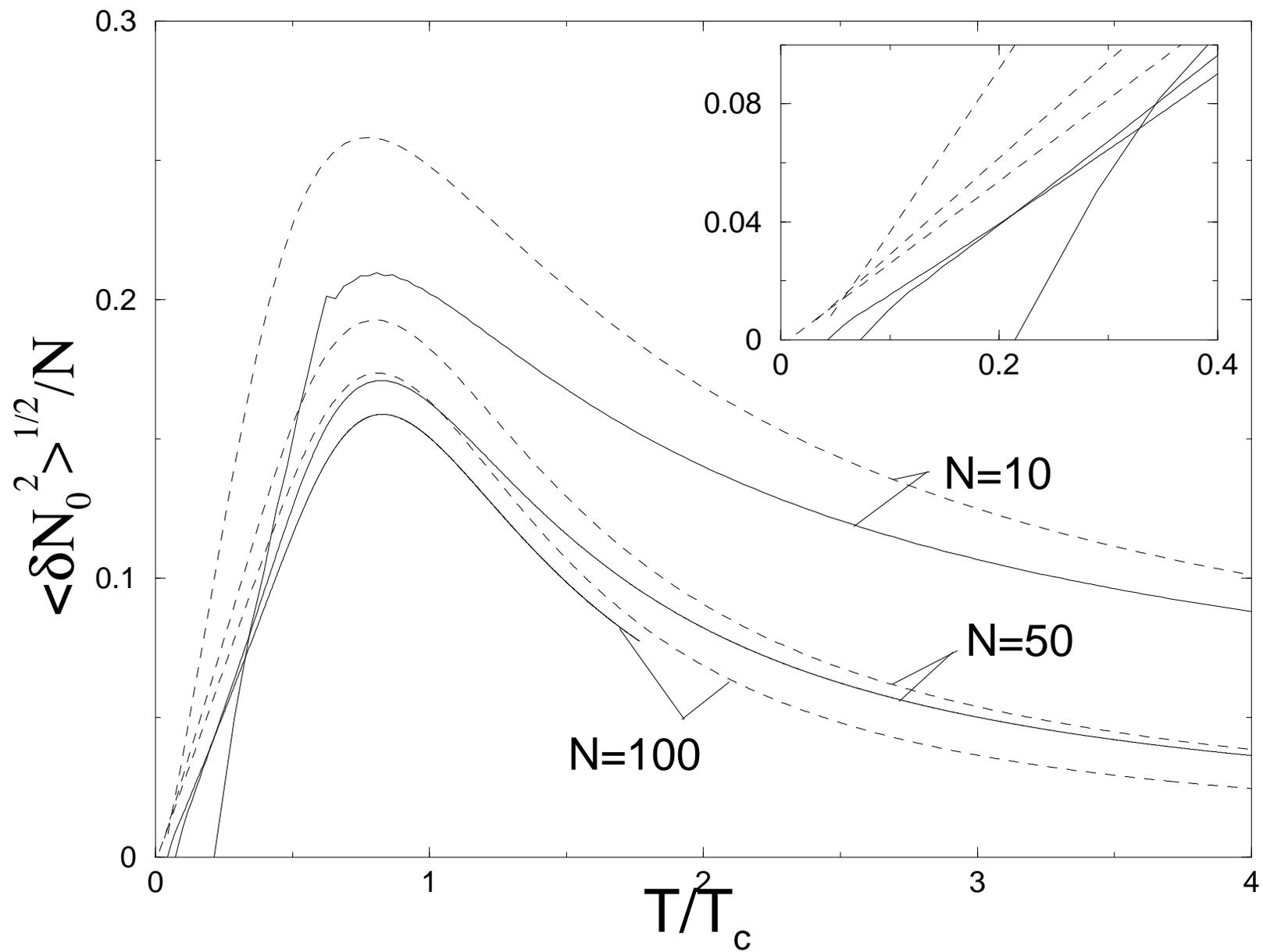
\end{document}